\documentclass{iaus}
\usepackage{graphicx}

\title[Evolution of dust in Pop III SNRs and its impact on 
Pop II.5 stars] 
{Evolution of newly formed dust in Population 
III supernova remnants and \\ its impact on the elemental composition 
of Population II.5 stars}

\author[Nozawa et al.]   
{Takaya Nozawa$^{1,2}$, Takashi Kozasa$^2$, Asao Habe$^2$, Eli Dwek$^3$, \\  
Hideyuki Umeda$^4$, Nozomu Tominaga$^5$, Keiichi Maeda$^{1,6}$, \and \\ 
Ken'ichi Nomoto$^{1,4,7}$}

\affiliation{
$^1$Institute for the Physics and Mathematics of the Universe, 
University of Tokyo, \\ Kashiwa, Chiba 277-8568, Japan \\
email: {\tt takaya.nozawa@ipmu.jp} \\[\affilskip]
$^2$Department of Cosmosciences, Graduate School of Science,
Hokkaido University, \\ Sapporo 060-0810, Japan \\ 
$^3$Laboratory for Astronomy and Solar Physics,
NASA Goddard Space Flight Center, \\ Greenbelt, MD 20771, USA \\
$^4$Department of Astronomy, School of Science,
University of Tokyo, \\ Bunkyo-ku, Tokyo 113-0033, Japan \\
$^5$Division of Optical and Infrared Astronomy, 
National Astronomical Observatory of Japan, \\ Mitaka, Tokyo 181-8588, 
Japan \\
$^6$Max-Planck-Institut f\"ur Astrophysik, 85741 Garching, Germany \\
$^7$Research Center for the Early Universe, School of Science,
University of Tokyo, \\ Bunkyo-ku, Tokyo 113-0033, Japan}

\pubyear{2008}
\volume{255}  
\pagerange{1-5}
\jname{Low-Metallicity Star Formation: From the First Stars to Dwarf Galaxies}
\editors{L.K. Hunt, S. Madden \& R. Schneider, eds.}
\begin{document}

\maketitle

\begin{abstract}
We investigate the evolution of dust formed in Population III supernovae 
(SNe) by considering its transport and processing by sputtering within 
the SN remnants (SNRs).
We find that the fates of dust grains within SNRs heavily depend on their 
initial radii $a_{\rm ini}$.
For Type II SNRs expanding into the ambient medium with density of 
$n_{\rm H,0} = 1$ cm$^{-3}$, grains of $a_{\rm ini} < 0.05$ $\mu$m are 
detained in the shocked hot gas and are completely destroyed, 
while grains of $a_{\rm ini} > 0.2$ $\mu$m are injected into the 
surrounding medium without being destroyed significantly.
Grains with $a_{\rm ini}$ = 0.05--0.2 $\mu$m are finally trapped in the 
dense shell behind the forward shock.
We show that the grains piled up in the dense shell enrich the gas up to
10$^{-6}$--10$^{-4}$ $Z_\odot$, high enough to form low-mass stars with 
0.1--1 $M_\odot$.
In addition, [Fe/H] in the dense shell ranges from $-6$ to $-4.5$, which 
is in good agreement with the ultra-metal-poor stars with [Fe/H] $< -4$.
We suggest that newly formed dust in a Population III SN can have great 
impacts on the stellar mass and elemental composition of Population II.5 
stars formed in the shell of the SNR.

\end{abstract}

\firstsection 
\section{Introduction}

The first dust in the universe plays critical roles in the subsequent 
formation processes of stars and galaxies.
Dust grains provide additional pathways for cooling of gas in metal-poor 
molecular clouds through their thermal emission and formation of H$_2$ 
molecules on the surface (e.g., Cazaux \& Spaans 2004). 
In particular, the presence of dust decreases the values of the critical 
metallicity to $10^{-6}$--$10^{-4}$ $Z_\odot$ (Omukai et al. 2005; 
Schneider et al. 2006; Tsuribe \& Omukai 2006), where the transition of
star formation mode from massive Population III stars to low-mass 
Population II stars occurs.
Since absorption and thermal emission by dust grains strongly depend on 
their composition, size distribution, and amount, it is essential to 
clarify the properties of dust in the early epoch of the universe,
in order to elucidate the evolutional history of stars and galaxies.

Dust grains at redshift $z > 5$ are considered to have been predominantly 
produced in supernovae (SNe).
Theoretical studies have predicted that dust grains of 0.1--2 $M_\odot$ 
and 10--60 $M_\odot$ are formed in the ejecta of primordial Type II SNe 
(SNe II, Todini \& Ferrara 2001; Nozawa et al. 2003) and 
pair-instability SNe (PISNe, Nozawa et al. 2003; Schneider et al. 2004), 
respectively.
However, the newly formed dust is reprocessed via sputtering in the hot 
gas swept up by the reverse and forward shocks that are generated from
the interaction between the SN ejecta and the surrounding medium.
Thus, the size and mass of the dust can be greatly modified before being 
injected into the interstellar medium (ISM, Bianchi \& Schneider 2007; 
Nozawa et al. 2007).

In this proceedings we present the results of the calculations for the 
evolution of newly formed dust within Population III SN remnants (SNRs), 
based on the dust formation model by Nozawa et al. (2003).
We investigate the transport of dust and its processing by sputtering 
in the shocked hot gas, and report the size and amount of dust injected 
from SNe into the ISM.
It is also shown that a part of the surviving dust grains are piled up in 
the dense SN shell formed behind the forward shock and can enrich the gas 
in the dense shell up to 10$^{-6}$--10$^{-4}$ $Z_\odot$.
We suppose that newly condensed dust in the SN ejecta has significant 
influences on the elemental abundances of Population II.5 stars, that is, 
the second-generation stars formed in the dense shell of Population III 
SNRs.

\section{Evolution of Dust in Population III SNRs}

We first describe the models of calculations for the evolution of dust 
within SNRs in short.
The time evolution of a SNR is numerically solved by assuming spherical 
symmetry.
We adopt the hydrodynamic models of Population III SNe II with the 
progenitor mass of $M_{\rm pr}$ = 13, 20, 25, and 30 $M_\odot$ and 
explosion energy of $10^{51}$ ergs by Umeda \& Nomoto (2002) for the 
initial condition of gas in the ejecta.
For the ambient medium, we consider the uniform medium with the hydrogen 
number density of $n_{\rm H, 0}=$ 0.1, 1, and 10 cm$^{-3}$.
For the model of dust in the He core, we adopt the dust grains formed in 
the unmixed ejecta by Nozawa et al. (2003).
Treating dust grains as test particles, we calculate the destruction and 
dynamics of dust by taking account into the size distribution as well as
the spatial mass distribution of each grain species.

\begin{figure}
\begin{center}
 \includegraphics[width=3.3in]{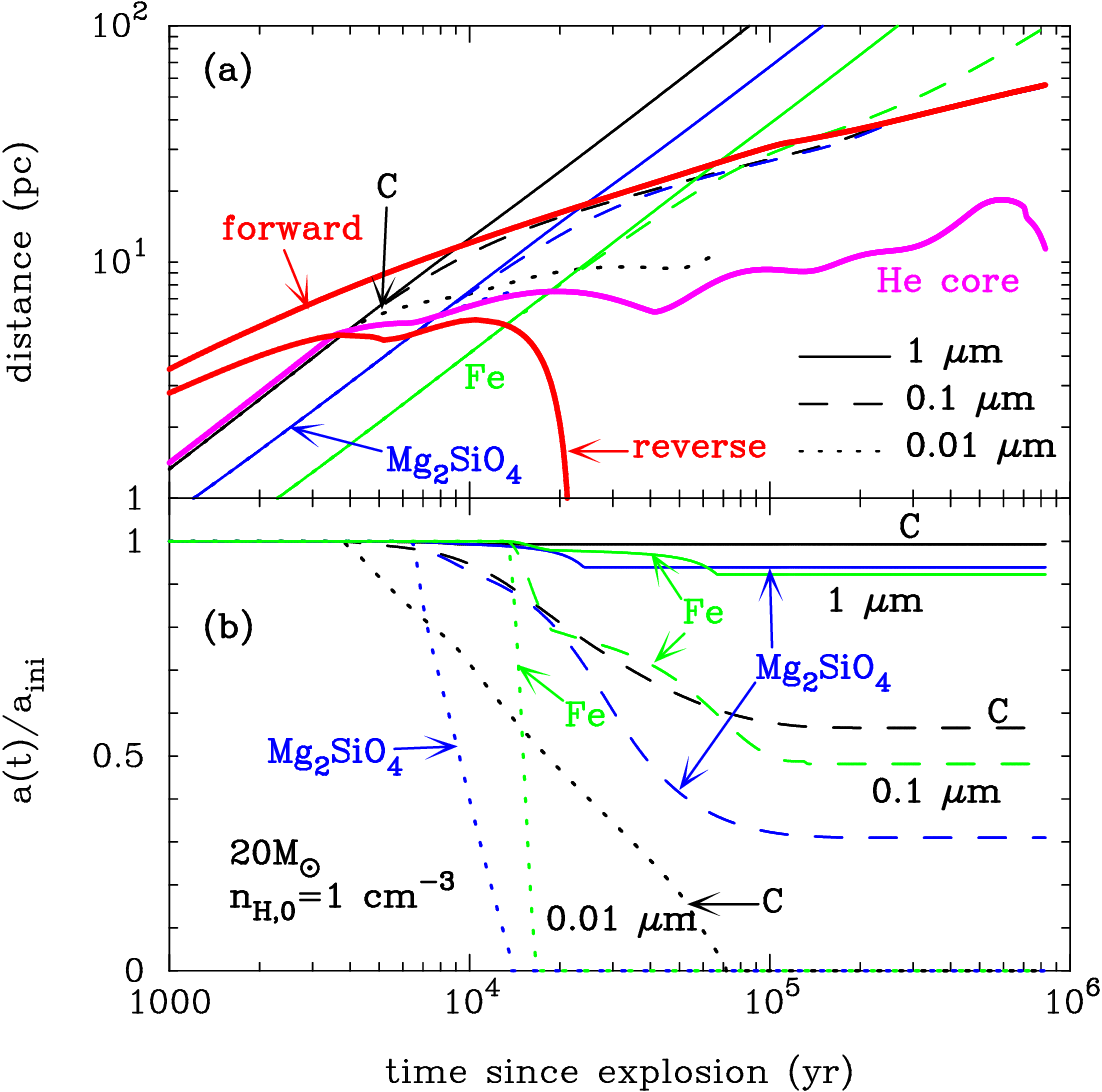} 
 \caption{
  (\textit{a}) Trajectories of C, Mg$_2$SiO$_4$, and Fe grains within the 
  SNR for $M_{\rm pr}=20$ $M_{\odot}$ and $n_{\rm H,0}=1$ cm$^{-3}$ and 
  (\textit{b}) the time evolutions of their radii relative to
  the initial ones. 
  The evolution of dust with $a_{\rm ini}=$ 0.01, 0.1, and 1 $\mu$m 
  is shown by the dotted, dashed, and solid lines, respectively.
  The thick solid lines in ({\it a}) indicate the positions of the forward 
  shock, the reverse shocks, and the surface of the He core.
}
   \label{fig1}
\end{center}
\end{figure}

The results of calculations are shown in Figure 1.
Figure 1{\it a} shows the trajectories of C, Mg$_2$SiO$_4$, and Fe grains 
within the SNR for $M_{\rm pr}=20$ $M_\odot$ and $n_{\rm H,0}=1$ 
cm$^{-3}$, and Figure 1{\it b} shows the time evolutions of their radii 
relative to the initial ones.
The positions of the forward shock, the reverse shock, and the surface of 
the He core are depicted by the thick solid lines in Figure 1{\it a}.
Initially, newly formed dust grains are expanding with the cool gas in the 
ejecta, and thus they undergo neither gas drag nor processing by sputtering.
However, once they intrude into the hot gas swept up by the reverse shock 
penetrating into the ejecta, they acquire the high velocities relative to 
the gas and are eroded by kinetic and/or thermal sputterings.
The evolutions of dust grains after colliding with the reverse shock 
heavily depend on their initial radii $a_{\rm ini}$ and compositions.

For example, C grains formed in the outermost region of the He core 
encounter the reverse shock at 3650 yr.
Since the deceleration by the gas drag is more efficient for smaller 
grains, C grains with $a_{\rm ini}=0.01$ $\mu$m quickly slow down.
These small grains are eventually trapped in the hot gas ($\ge 10^6$ K) 
generated from the passage of the reverse and forward shocks and are 
completely destroyed by thermal sputtering.
C grains of $a_{\rm ini} = 0.1$ $\mu$m reduce their sizes by sputtering 
but cannot be completely destroyed.
These grains are finally captured in the dense SN shell formed behind the 
forward shock at $\sim$2$\times$10$^5$ yr, where the gas temperature is 
too low ($<$ $10^5$ K) to erode the dust grains by thermal sputtering.
C grains with $a_{\rm ini}=1$ $\mu$m, which are not efficiently 
decelerated by the gas drag, can pass through the forward shock front 
and are injected into the ambient medium without being destroyed 
significantly.

Mg$_2$SiO$_4$ grains, which are formed in the O-rich layer, collide with 
the reverse shock at about 6000 yr, but the dependence of their 
evolutions on the initial radius is the same as C grains.
On the other hand, Fe grains formed in the innermost region of the ejecta 
hit the reverse shock after 13000 yr, and its 0.1 $\mu$m-sized grains are 
injected into the ambient medium because of the high bulk density.
 
As is shown above, the small dust grains formed in the SN ejecta are 
predominantly destroyed by sputtering within the SNR.
Thus, the size distribution of the surviving dust is dominated by
larger grains, compared to that at its formation.
Note that the evolution of dust within SNRs and thus the resulting size 
distribution of dust do not depend on the progenitor mass considered here,
because their explosion energies are the same and the time evolutions of 
the gas temperature and density within SNRs are similar.
On the other hand, the ambient gas density strongly affects the 
evolution of dust in SNRs.
The higher density in the ambient medium results in the higher density of
the shocked gas and causes the efficient erosion and deceleration of dust
due to more frequent collisions with the hot gas.
Therefore, the initial radius below which dust is completely destroyed 
increases with increasing the ambient gas density and is 0.01, 0.05, and 
0.2 $\mu$m for $n_{\rm H, 0} =$ 0.1, 1, and 10 cm$^{-3}$, respectively.
As a result, the total mass of the surviving dust is smaller for the
higher ambient density and ranges from 0.01 to 0.8 $M_\odot$ for 
$n_{\rm H, 0}=$ 10 to 0.1 cm$^{-3}$, depending on the size distribution 
of dust formed in each SN.

\begin{table}
\begin{center}
  \caption{Metallicities, [Fe/H], and abundances of C, O, Mg, and Si 
relative to Fe in the dense shell of primordial SN II remnants 
for various ambient gas densities.}
  \label{tab1}
\begin{tabular}{lcccccc}

\hline
  {$M_{\rm pr}$ ($M_{\odot}$)}~~ & 
~~{$\log(Z/Z_\odot)$}~~ & ~~{[Fe/H]}~~ & 
~~{[C/Fe]}~~ & ~~{[O/Fe]}~~ & ~~{[Mg/Fe]}~~ & ~~{[Si/Fe]}  \\
\hline
\multicolumn{7}{c}{$n_{\rm H,0}=0.1$ cm$^{-3}$} \\
\hline
13 & $-5.89$ & $-6.43$ & $-0.274$ & $-0.699$ & $-0.230$ & $1.92$ \\
20 & $-5.44$ & $-5.20$ & $0.117$  & $-0.595$ & $0.034$  & $0.410$ \\
25 & $-5.55$ & $-5.90$ & $1.11$   & $-1.42$  & $-0.500$ & $-0.552$ \\
30 & $-5.33$ & $-5.56$ & $0.566$  & $-0.043$ & $0.739$  & $0.866$ \\
\hline
\hline
\multicolumn{7}{c}{$n_{\rm H,0}=1$ cm$^{-3}$} \\
\hline
13 & $-4.72$ & $-5.15$ & $1.11$  & $-0.555$ & $-0.459$ & $1.01$ \\
20 & $-4.68$ & $-5.53$ & $0.992$ & $0.585$  & $1.16$   & $1.87$ \\
25 & $-4.79$ & $-5.23$ & $1.09$  & $-0.412$ & $0.407$  & $0.989$ \\
30 & $-4.60$ & $-5.11$ & $0.797$ & $0.242$  & $1.09$   & $1.26$  \\
\hline
\hline
\multicolumn{7}{c}{$n_{\rm H,0}=10$ cm$^{-3}$} \\
\hline
13 & $-4.40$ & $-4.13$ & $0.284$  & $-2.54$ & $-3.89$ & $0.599$ \\
20 & $-4.09$ & $-4.92$ & $0.946$  & $-2.15$ & $-1.80$ & $2.14$ \\
25 & $-3.91$ & $-5.10$ & $1.60$   & $0.122$ & $0.232$ & $2.34$ \\
30 & $-3.84$ & $-5.11$ & $-0.207$ & $0.375$ & $-1.23$ & $2.66$ \\
\end{tabular}
\end{center}
\end{table}

\section{Metallicities and Elemental Abundances of Population II.5 Stars}
 
In this section we discuss the influence of dust on the elemental 
composition of Population II.5 stars that are expected to form in the 
dense shell of Population III SNRs (Mackey et al. 2003; Salvaterra et al. 
2004; Machida et al. 2005).
As shown in the last section, the dust grains surviving the destruction 
but not injected into the ISM are piled up in the dense SN shell after 
10$^5$--10$^6$ yr.
This implies that the elemental composition of these piled-up grains can 
play an important role in the elemental abundances of Population II.5 
stars.
Furthermore, the existence of dust in the shell may enable the formation 
of stars with solar mass scales through its thermal emission if the gas 
is enriched to the critical metallicities (Omukai et al. 2005; 
Schneider et al. 2006; Tsuribe \& Omukai 2006).
Thus, we calculate the metallicities and metal abundances in the dense 
shell based on the elemental composition of piled-up grains, and compare 
with the observed abundance patterns of low-mass hyper-metal-poor (HMP) 
and ultra-metal-poor (UMP) stars.

Table 1 summarizes the calculated metallicities and elemental abundances 
in the dense shell of SN II remnants as a function of the ambient gas 
density.
It should be noted that metallicity in the shell ranges from 10$^{-6}$ 
to 10$^{-4}$ $Z_\odot$, which is considered to cause the formation of 
stars with 0.1--1 $M_\odot$.
In addition, most of calculated [Fe/H] are in the range from $-6$ to 
$-4.5$, which is in good agreement with those for HMP and UMP stars.
We also plot in Figure 2a the abundances of C, O, Mg, and Si relative 
to Fe in the shell for $n_{\rm H,0} = 0.1$ and 1 cm$^{-3}$.
We can see that the calculated abundances of Mg and Si showing 1--100 
times overabundances are consistent with the observations of HMP and UMP 
stars.
Because the elemental composition of dust piled up in the shell can 
reproduce the abundance patterns of Fe, Mg, and Si in HMP and UMP stars, 
it is considered that the transport of dust separated from metal-rich gas 
within SNRs can be attributed to the elemental compositions of HMP and 
UMP stars, if they are Population II.5 stars.

\begin{figure}
\begin{center}
 \includegraphics[width=2.6in]{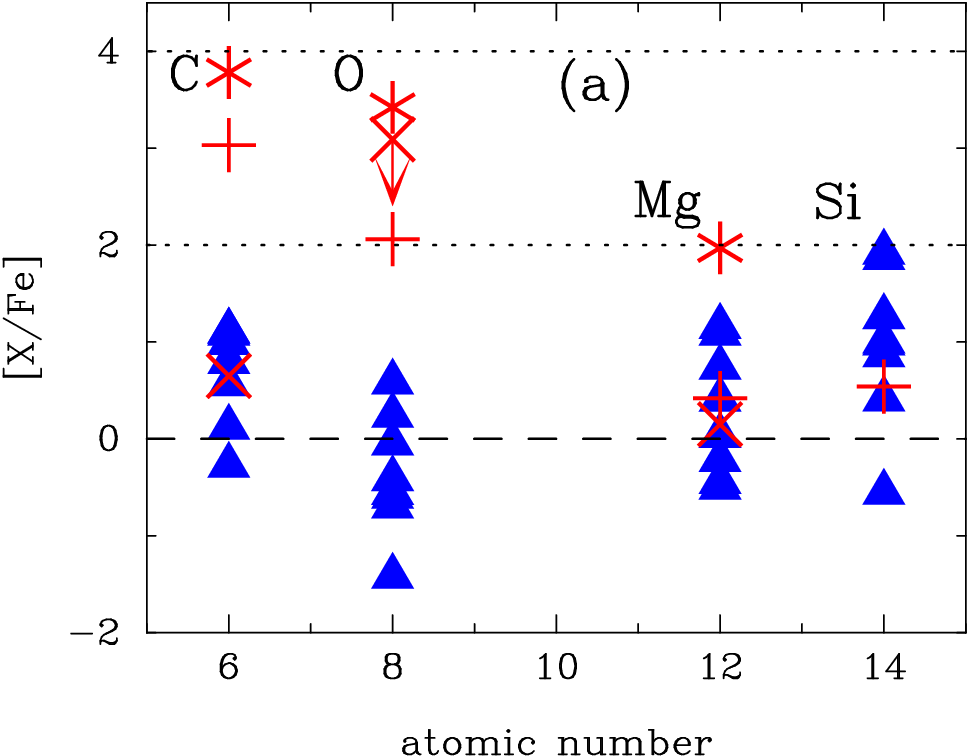} 
 \includegraphics[width=2.6in]{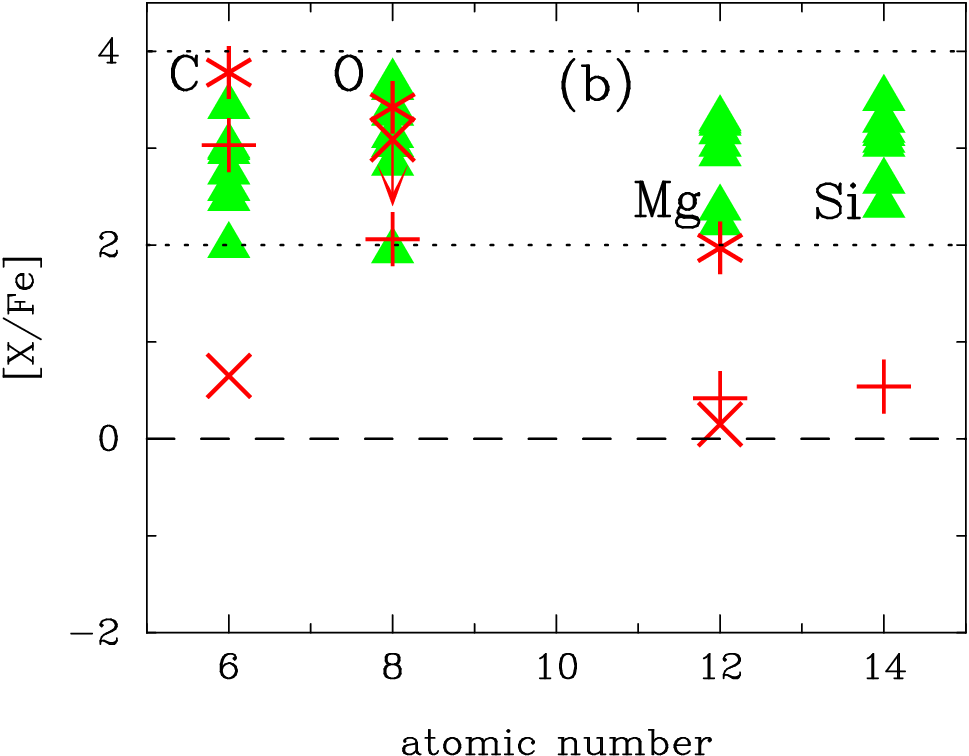} 
 \caption{
  Abundances of C, O, Mg and Si relative to Fe in the dense shell of SN
  II remnants for $n_{\rm H,0} = 0.1$ and 1 cm$^{-3}$ ({\it filled 
  triangles}); (\textit{a}) derived from the elemental composition of the 
  grains piled up in the shell, and (\textit{b}) derived from the elemental 
  composition of the piled-up grains and the gas outside the innermost 
  Fe layer.
  For observational data of HMP and UMP stars, the 3-D corrected 
  abundances are adopted and are denoted by plus (HE0107-5240 with 
  [Fe/H] = -5.62, Collet et al. 2006), asterisk (HE1327-2326 with 
  [Fe/H] = -5.96, Frebel et al. 2008), and cross (HE0557-4840 with
  [Fe/H] = -4.75, Norris et al. 2007). 
}
   \label{fig2}
\end{center}
\end{figure}

However, as can be seen from Figure 2${\it a}$, no model considered here
can reproduce $10^2$--$10^4$ times excesses of C and O observed in HMP 
stars.
One of the reasons is that in the calculation we assumed that the 
metal-rich gas in the SN ejecta does not mix with the gas in the shell.
Then we examine the abundance patterns in the shell by assuming that 
besides the piled-up grains, the gas outside the innermost Fe layer in the 
ejecta is incorporated into the shell. 
The results are shown in Figure 2{\it b}.
In this case we can reproduce the very large overabundances of C and O, but 
the excesses of Mg and Si are too large ($\ge$100 times) to agree with the 
observations.
However, it could be possible to reproduce the abundance patterns of 
refractory elements observed in HMP stars unless the Si-Mg-rich layer is 
mixed into the shell.
Unfortunately, it has been still debated what extent of the gas in the 
ejecta can mix into the SN shell when Population II.5 stars form.
Nevertheless, we can conclude that newly formed dust in a Population III 
SN can have great impacts on the stellar mass and metal abundance of 
Population II.5 stars, if the metal-rich gas is not significantly 
incorporated into the dense gas shell. \\

 This work has been supported in part by World Premier International 
 Research Center Initiative (WPI initiative), MEXT, Japan and by the
 Grant-in-Aid for Scientific Research of the Japan Society for the 
 Promotion of Science (18104003, 19740094).

\end{document}